\documentclass[journal=jacsat,manuscript=article]{achemso}

\usepackage{chemformula} 
\usepackage[T1]{fontenc} 
\usepackage{hyperref}
\usepackage[version=3]{mhchem}

\usepackage{siunitx}

\author{Jem Pitcairn}
\affiliation[Nottingham]{School of Chemistry, University of Nottingham,
University Park, Nottingham, NG7 2RD, United Kingdom}

\author{Andrea Iliceto}
\affiliation[Birmingham]{School of Metallurgy and Materials, University
of Birmingham, Elms Rd, Edgbaston, Birmingham B15 2TT, United Kingdom}

\author{Laura Cañadillas-Delgado}
\affiliation[Institut Laue-Langevin]{Institut Laue-Langevin, 71 avenue
des Martyrs - CS 20156, 38042 Grenoble, France}

\author{Oscar Fabelo}
\affiliation[Institut Laue-Langevin]{Institut Laue-Langevin, 71 avenue
des Martyrs - CS 20156, 38042 Grenoble, France}

\author{Cheng Liu}
\affiliation[Cambridge]{Cavendish Laboratory, Department of Physics,
University of Cambridge, JJ Thomson Avenue, Cambridge CB3 0HE, United
Kingdom}

\author{Christian Balz}
\affiliation[ISIS]{ISIS Neutron and Muon Source, STFC Rutherford
Appleton Laboratory, Harwell Oxford, Didcot OX11 0QX, United Kingdom}

\author{Andreas Weilhard}
\affiliation[Nottingham]{School of Chemistry, University of Nottingham,
University Park, Nottingham, NG7 2RD, United Kingdom}

\author{Stephen P. Argent}
\affiliation[Nottingham]{School of Chemistry, University of Nottingham,
University Park, Nottingham, NG7 2RD, United Kingdom}

\author{Andrew J. Morris}
\affiliation[Birmingham]{School of Metallurgy and Materials, University
of Birmingham, Elms Rd, Edgbaston, Birmingham B15 2TT, United Kingdom}

\author{Matthew J. Cliffe}
\affiliation[Nottingham]{School of Chemistry, University of Nottingham,
University Park, Nottingham, NG7 2RD, United Kingdom}

\email{matthew.cliffe@nottingham.ac.uk}

\title[Low D MOM for S=2 Haldane]
  {Low-dimensional metal-organic magnets as a route towards the \(S=2\)
Haldane phase}

\abbreviations{IR,NMR,UV}
\keywords{American Chemical Society, \LaTeX}

\begin{document}

\begin{abstract}
  Metal-organic magnets (MOMs), modular magnetic materials where metal
atoms are connected by organic linkers, are promising candidates for
next-generation quantum technologies. MOMs readily form low-dimensional
structures, and so are ideal systems to realise physical examples of key
quantum models, including the Haldane phase, where a topological
excitation gap occurs in integer-spin antiferromagnetic (AFM) chains.
Thus far the Haldane phase has only been identified for \(S=1\), with
\(S \geq 2\) still unrealised because the larger spin imposes more
stringent requirements on the magnetic interactions. Here, we report the
structure and magnetic properties of \ce{CrCl2(pym)} (pym=pyrimidine), a
new quasi-1D \(S=2\) AFM MOM. We show, using X-ray and neutron
diffraction, bulk property measurements, density-functional theory
calculations and inelastic neutron spectroscopy (INS) that
\ce{CrCl2(pym)} consists of AFM \ce{CrCl2} spin chains
(\(J_1=-1.13(4)\;\)meV) which are weakly ferromagnetically coupled
through bridging pym (\(J_2=0.10(2)\;\)meV), with easy-axis anisotropy
(\(D=-0.15(3)\;\)meV). We find that although small compared to \(J_1\),
these additional interactions are sufficient to prevent observation of
the Haldane phase in this material. Nevertheless, the proximity to the
Haldane phase together with the modularity of MOMs suggests that layered
Cr(II) MOMs are a promising family to search for the elusive \(S=2\)
Haldane phase.
\end{abstract}

\section{Introduction}

\begin{figure}
\hypertarget{fig:cryst}{%
\centering
\includegraphics{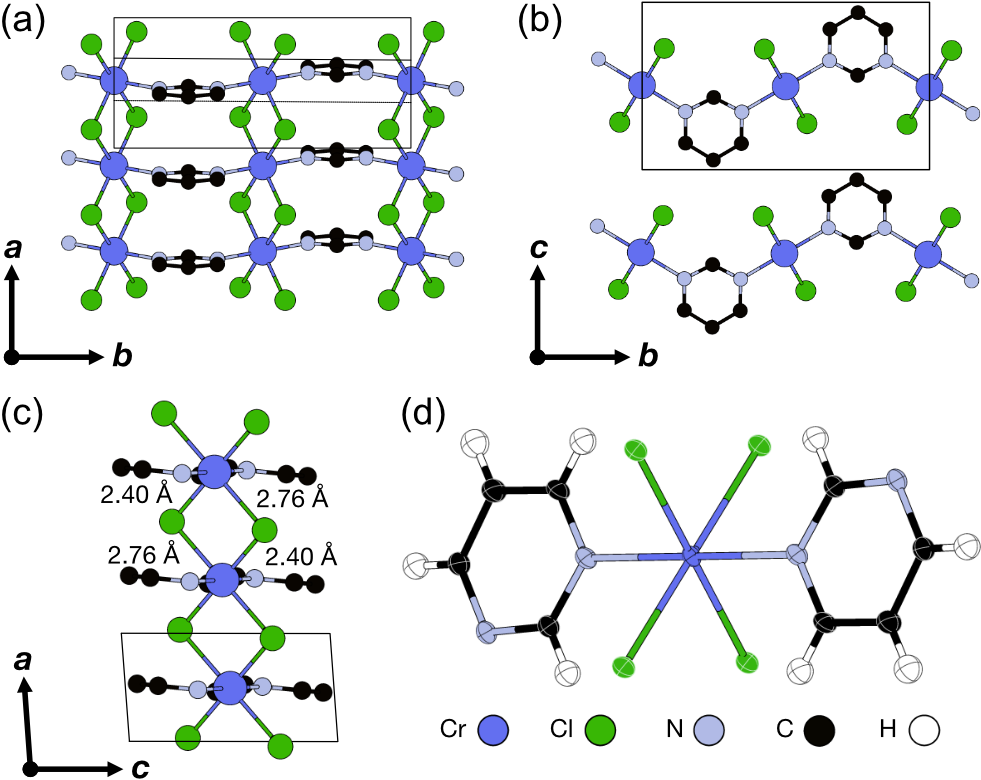}
\caption{Crystal structure of \ce{CrCl2(pym)} viewed along the (a)
\emph{c}, (b) \emph{a} and (c) \emph{b} axes. Cr--Cl bond lengths are
labelled and H atoms are omitted for clarity. (d) ORTEP diagram showing
the coordination environment.}\label{fig:cryst}
}
\end{figure}

Metal-organic magnets (MOM) are assembled from metal nodes bridged by
organic molecular linkers into extended
networks.\citep{thorarinsdottirMetalOrganicFramework2020} This gives
them a number of advantages over conventional inorganic magnets: there
is a much wider diversity of organic than atomic
ligands;\citep{zhaoTuningTopologyFunctionality2011} the modularity of
their construction allows for tuning of interactions while retaining the
topology\citep{cortijoModulationMagneticProperties2013} and their longer
lengths facilitate magnetic
low-dimensionality,\citep{canepaWhenMetalOrganic2013, harcombeOnedimensionalMagneticOrder2016}
and thus enhanced quantum
fluctuations.\citep{vasilievMilestonesLowDQuantum2018} Perhaps most
excitingly, it has recently been demonstrated that redox-active radical
ligands can introduce into MOFs both high electronic conductivity (0.45
Scm\(^{-1}\))\citep{ziebelControlElectronicStructure2018} and strong
magnetic
interactions,\citep{pedersenFormationLayeredConductive2018, perlepeMetalorganicMagnetsLarge2020}
despite the long-distances between metal centres. This suggests that
MOMs could form the basis for practical new quantum
technology.\citep{murphyExchangeBiasLayered2021, parkMagneticOrderingItinerant2021, yamadaDesigningKitaevSpin2017, huangSuperconductivityCopperII2018a, coronadoMolecularMagnetismChemical2020}

MOM spin chains are now well established as host materials for
distinctively quantum behaviour, from spin fractionalisation in
\ce{Cu(C6H5COO)2.3H2O}\citep{denderDirectObservationFieldInduced1997} to
the quantum sine-Gordon physics of
Cu(pym)(NO\(_3\))(H\(_2\)O)\(_2\)\citep{zvyaginExcitationHierarchyQuantum2004}
and
\ce{[Cu(pym)(H2O)4]SiF6.H2O}.\citep{liuUnconventionalFieldInducedSpin2019}
One of the most striking quantum discoveries in MOMs was the measurement
of the topological Haldane gap in the antiferromagnetic \(S=1\) spin
chain MOM \ce{Ni(C2H8N2).2NO2(ClO4)}
\citep{renardPresumptionQuantumEnergy1987, haldaneNonlinearFieldTheory1983, haldaneContinuumDynamics1D1983}
and subsequent efforts have uncovered a number of other high-quality
model
systems.\citep{mutkaSupportHaldaneConjecture1989, takeuchiMagnetizationProcessHaldane1992, landeeLowtemperatureCrystalStructures1997, williamsNearidealMoleculebasedHaldane2020}
The Haldane phase is yet to be experimentally realised for spins
\(S>1\).

The difficulty of reaching the Haldane phase for \(S \geq 2\) is largely
because the size of the Haldane gap relative to the intrachain exchange,
\(\Delta / J_1\), decreases significantly from \(\Delta/J_1 = 0.41\) for
\(S=1\) to \(\Delta/ J_1 = 0.087\) for \(S=2\), making the gap both more
sensitive to the presence of single-ion anisotropy and non-Heisenberg
exchange interactions, and harder to detect when
present.\citep{schollwockHaldaneGapHidden1995} These challenges have
meant that although AFM \(S=2\) spin chains which could be candidates to
host the Haldane phase have been identified, the \(S=2\) gap has not yet
been
observed.\citep{stoneQuasionedimensionalSpinWaves2013, granrothNeutronscatteringStudiesAntiferromagnetic2002, leoneNeutronDiffractionStudy2004, stockOneDimensionalMagneticFluctuations2009, birkMagneticPropertiesManganese2011}
The combination of modularity and low-dimensionality of MOMs means they
are an ideal platform to search for the \(S=2\) Haldane phases. However,
the most synthetically accessible \(S=2\) transition metal ion is
\ce{Fe^{2+}}, which typically possesses large single ion anisotropy due
to its partially quenched \(^5T_{2\mathrm{g}}\) ground state, and other
\(S=2\) ions, \ce{Mn^{3+}} and \ce{Cr^{2+}}, are usually sensitive to
reduction or oxidation in ambient conditions. As a result, the chemistry
of MOMs which could host \(S=2\) Haldane phases is comparatively
underexplored, and their quantum states thus unrealised.

Here we report \ce{CrCl2(pym)}, a new 2D layered magnetic coordination
polymer consisting of \ce{CrCl2} chains bridged by pym ligands with a
structure analogous to that of the other transition metal monopyrimidine
chlorides (\ce{MCl2(pym)}, M = Mn, Fe, Co, Ni,
Cu),\citep{ferraroTransitionMetalII1969} the Mn, Co and Cu analogues of
which are reported to possess antiferromagnetic coupling without order
down to \(1.8\;\)K.\citep{zusaiMagnetismPyrimidineBridgedMetal2000} We
first describe its synthesis and structural characterisation using X-ray
diffraction, where the presence of a pronounced Jahn-Teller (JT)
distortion confirms the presence of \ce{Cr^{2+}}. We then go on to show
using comprehensive magnetic characterisation, including bulk
magnetisation, heat capacity measurements, powder neutron diffraction
(PND) and powder inelastic neutron scattering (INS) measurements of
fully protonated samples, that \ce{CrCl2(pym)} orders into a Néel ground
state at \(T_\mathrm{N} = 20.0(3)\) K, with AFM ordering along the
\ce{CrCl2} chain, FM coupling of the chains through pym and interlayer
FM correlations. Through detailed analysis of the neutron scattering
data, in combination with density-functional theory (DFT) calculations,
we quantitatively determine the size of the key magnetic interactions,
which suggest that \ce{CrCl2(pym)} is a well separated \(S=2\) AFM with
near isotropic single ion properties. We therefore suggest that through
careful ligand choice this family of MOMs offers a potential route to
realise the Haldane phase for \(S=2\).

\hypertarget{results}{%
\section{Results}\label{results}}

\hypertarget{synthesis-and-structure}{%
\subsection{Synthesis and structure}\label{synthesis-and-structure}}

We synthesised \ce{CrCl2(pym)} by reacting \ce{CrCl2} with pyrimidine.
We found that the monopyrimidine \ce{CrCl2(pym)} forms in a wide variety
of solvents and stoichiometries, and even \emph{via} neat combination
and with excess ligand, although bispyrimidine analogues are known for
other transition
metals.\citep{feyerhermWeakFerromagnetismVery2004, hashizumeDichlorobisPyrimidineNCobalt1999, kreitlowPressureDependenceC4N2H4mediated2005}
Single crystals suitable for X-ray diffraction measurements were grown
through vapour diffusion. We solved the structure from single-crystal
X-ray diffraction (SCXRD) data and found that \ce{CrCl2(pym)}
crystallises in the monoclinic space group \(P2_1/m\) with two formula
units in the unit cell (Tab. S1). The \ce{Cr^{2+}} are coordinated by
four \ce{Cl^{-}} ligands and two \ce{N} atoms from the pyrimidine
ligands, which forms a distorted \ce{CrCl4N2} octahedron (Fig.
\ref{fig:cryst}c \& d). The chromium octahedra edge-share through the
\ce{Cl^-} ligands along the crystallographic \emph{a} direction, and
these chains are connected by pyrimidine ligands along the
crystallographic \emph{b} direction with an alternating orientation to
form corrugated layers (Fig. \ref{fig:cryst}a). These layers stack in
the crystallographic \emph{c} direction through van der Waals
interactions (Fig. \ref{fig:cryst}b). The \ce{Cr^{2+}} ion has a large
JT distortion, with a long \ce{Cr-Cl} bond length of
\(d_\mathrm{Cr-Cl} = 2.761(5)\;\)Å, comparable to the complex
\ce{\ce{Cr^{2+}}Cl2(pyridine)4} \(d_\mathrm{Cr-Cl} = 2.803(1)\)
Å,\citep{cottonExperimentalTheoreticalStudy1995} confirming the
\ce{Cr^{2+}} oxidation state. Powder X-ray diffraction performed after
exposure to air for one month show the lattice distortion resulting from
this JT distortion is retained, demonstrating that the bulk of the
sample maintains the \ce{Cr^{2+}} oxidation state after exposure to air
(Fig. S4).

\hypertarget{magnetic-susceptibility}{%
\subsection{Magnetic susceptibility}\label{magnetic-susceptibility}}

\begin{figure}
\hypertarget{fig:chi}{%
\centering
\includegraphics{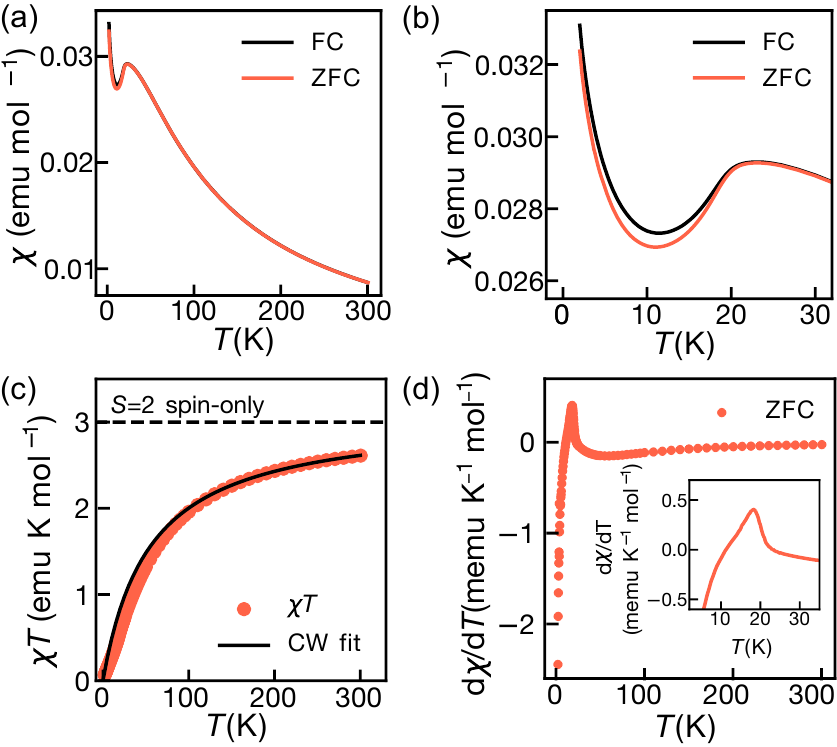}
\caption{Magnetic susceptibility, \(\chi\), measurements of
\ce{CrCl2(pym)}. (a) \(\chi(T)\) measured in zero-field cooled (ZFC) and
field cooled (FC) conditions from 2--300 K. (b) \(\chi(T)\) data
highlighted for 2--30 K. (c) \(\chi T (T)\) in ZFC and FC conditions
2--300 K, with Curie-Weiss fit carried out over \(300>T>150\;\)K. Dashed
line shows the \(S=2\) spin-only limit. (d) ZFC \(\frac{d \chi}{dT}(T)\)
over 2--300 K. Inset: ZFC \(\frac{d \chi}{dT}(T)\) over 2--35
K.}\label{fig:chi}
}
\end{figure}

As we expected \ce{CrCl2(pym)} to be an \(S=2\) 2D magnet, we measured
its temperature dependent magnetic susceptibility, \(\chi(T)\). The
sample was measured under field cooled (FC) and zero-field cooled (ZFC)
conditions in a 0.01 T \emph{dc} field from 2 K to 300 K. These data
show a broad peak at 20-25 K characteristic of short-range ordering and
low-dimensional magnetism (Fig. \ref{fig:chi}a). The
\(\frac{d \chi}{dT} (T)\) data show a discontinuity at 20 K, indicating
a phase transition from a disordered magnetic state to a long-range
ordered AFM state (Fig. \ref{fig:chi}d). Fitting \(\chi^{-1}(T)\) data
to the Curie-Weiss law gave a Curie constant, \(C = 3.08(1)\) emu K
mol\(^{-1}\), in good agreement with the presence of high-spin
\ce{Cr^{2+}} (\(C = 3\) emu K mol\(^{-1}\)) (Fig. \ref{fig:chi}c \& d).
The Curie-Weiss temperature is significant and negative,
\(\theta=-54.1(5)\) K, indicating net antiferromagnetic interactions
(Fig. \ref{fig:chi}d), and isothermal magnetisation measurements carried
out at 2 K show that saturation is not reached at fields of 5 T (Fig.
S7). While \(M(H)\) is linear in \(\mu _0H>1\) T, there is a small
sigmoid feature at \(\mu _0H<1\) T consistent with minor paramagnetic
impurities.

The rise in \(\chi(T)\) below \(T=10\) K indicates the presence of small
quantities of paramagnetic spins, which we determined to be 1.1(1) spin
\% from fitting of the Curie-like tail (Fig.
S13).\citep{johnstonUnifiedMolecularField2015} This Curie-like tail may
be caused free-spins at chain-ends or \ce{Cr^{3+}} formed due to surface
oxidation (Fig. \ref{fig:chi}b). Indeed, measurement of the magnetic
susceptibility of \ce{CrCl2(pym)} after air exposure showed a large
increase in the paramagnetic contribution, 15.0(2) spin \% (Fig. S6),
and X-ray photoelectron spectroscopy (XPS) of this air-exposed sample
primarily detected oxidised Cr (Fig. S8), with \ce{Cr^{3+}},
\ce{Cr^{6+}} and metallic Cr present, as well as O 1s peaks consistent
with the formation of
\ce{Cr(OH)3}.\citep{gazzoliChromiumOxidationStates1992}

\hypertarget{heat-capacity}{%
\subsection{Heat capacity}\label{heat-capacity}}

\begin{figure}
\hypertarget{fig:HC}{%
\centering
\includegraphics{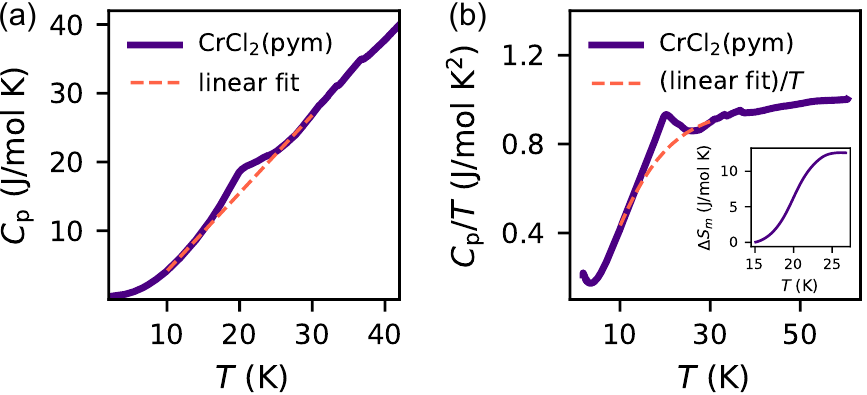}
\caption{(a) Heat capacity as a function of temperature,
\(C_{\mathrm{p}}(T)\), with the non-magnetic background approximated by
a linear fit over the region 10 to 30 K (dashed). (b)
\(C_{\mathrm{p}}/T(T)\), with non-magnetic background (dashed). Inset:
entropy near \(T_\mathrm{N}\).}\label{fig:HC}
}
\end{figure}

The molar heat capacity, \(C_{\mathrm{p}}(T)\), of \ce{CrCl2(pym)} was
measured from 2--60 K. We found a peak in \(C_{\mathrm{p}}(T)\) occurred
at 20.0(3) K (Fig. \ref{fig:HC}a), consistent with the magnetic phase
transition observed in the magnetic susceptibility data (Fig.
\ref{fig:chi}a). We obtained an estimate of the entropy of magnetic
ordering by integrating \(C_P/T(T)\) after subtraction of a linear
background (\(10-15\) K and \(27-30\) K)(Fig. \ref{fig:HC}b), to account
for phononic contributions. We found that the measured value of magnetic
entropy (\(S_\mathrm{exp.} = 12.7(4)\) Jmol\(^{-1}\)K\(^{-1}\)) is
slightly reduced from the expected value (\(S_\mathrm{calc.} = 13.4\)
Jmol\(^{-1}\)K\(^{-1}\)). The small features present in the data between
30--40 K are due to instrumental error.

\hypertarget{neutron-diffraction}{%
\subsection{Neutron Diffraction}\label{neutron-diffraction}}

\begin{figure}[h]

\centering

\includegraphics{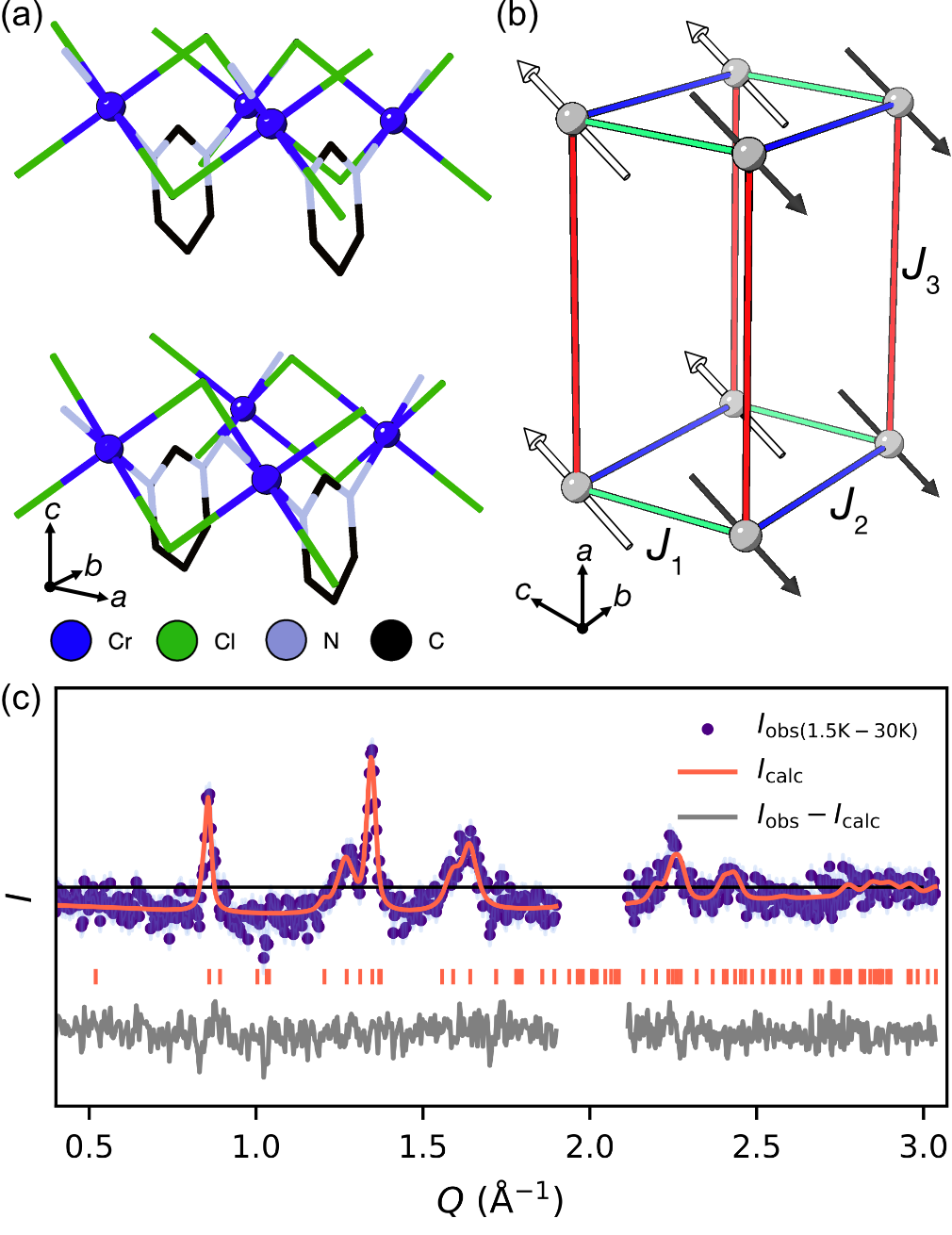}

 \caption{(a) The crystal structure of \ce{CrCl2(pym)}, nuclear axes shown. (b) The magnetic structure, highlighting the three most important exchange interactions, $J_n$, magnetic axes shown. (c) Rietveld refinement of temperature subtracted neutron scattering data. Data between $Q=1.9$ Å$^{-1}$ and $2.1$ Å$^{-1}$ were excluded from the refinement due to incomplete subtraction of nuclear Bragg peaks due to thermal expansion.}

\label{fig:pndmag}

\end{figure}

Our bulk measurements thus strongly suggested the presence of long-range
magnetic order. To determine the nature of this magnetic ground state we
carried out PND using instrument D1B at the ILL on a 5 g non-deuterated
sample of \ce{CrCl2(pym)}. We measured the neutron diffraction pattern
at two temperatures: \(T=1.5\;\)K below \(T_\mathrm{N}\), and
\(T=30\;\)K above. We isolated the magnetic scattering from instrumental
background and nuclear scattering contributions by subtracting the high
temperature dataset from the low temperature dataset (Fig.
\ref{fig:pndmag}c), which allowed us to identify the magnetic Bragg
peaks. We were able to index these reflections with a propagation vector
\(\textbf{k}= \frac{1}{2}00\) and using symmetry-mode analysis in the
ISODISTORT software suite\citep{campbellISODISPLACEWebbasedTool2006} we
identified there were two possible irreducible representations (irreps),
\(mY^-_1\) and \(mY^-_2\) in Miller and Love's
notation.\citep{cracknellKroneckerProductTables1979} After calibration
of the nuclear scale factor through Rietveld refinement of nuclear
structure against the high temperature dataset, we carried out Rietveld
refinement of the magnetic structure using each irrep against the
temperature subtracted dataset. We found for both nuclear and magnetic
refinement that a \(hkl\)-dependent peak broadening term was necessary
to account for the variation in measured peak widths. This showed that
only the \(mY^-_1\) irrep was consistent with experimental data (Fig.
\ref{fig:pndmag}c). The \(mY^-_1\) irrep lowers the symmetry of the
structure to \(P_c2_1/c\) with the magnetic unit cell relating to the
nuclear cell as follows: \(a_\mathrm{mag.}=c_\mathrm{nuc.}\),
\(b_\mathrm{mag.}=b_\mathrm{nuc.}\) and
\(c_\mathrm{mag.}=2a_\mathrm{nuc.}\) (Fig. \ref{fig:pndmag}a \& b).

The magnetic structure derived from this refinement is a collinear
structure consisting of antiferromagnetically correlated \ce{CrCl2} spin
chains ferromagnetically correlated through the pym ligands, with
interlayer ferromagnetic correlations (Fig. \ref{fig:pndmag}b). The
refined magnetic moment was Cr was determined to be
\(M_0=2.61(7) \mu_{\mathrm{B}}\), significantly less than the spin-only
value of \(M = gS = 4\,\mu_{\mathrm{B}}\).

The magnetic moments in our model lie within the \emph{ac}-plane,
however components along the \emph{b}-direction would be permitted by
symmetry. The presence of a component along \emph{b} would result in
intensity at the \(011_\mathrm{mag.}\) peak position (\(Q=1.00\)
Å\(^{-1}\)) which is not seen in our data and so any non-collinearity
must be small, \(\theta<8^{\circ}\). The background of this subtracted
\(I_{\mathrm{1.5 K}}-I_{\mathrm{30 K}}\) dataset contains a broad
negative feature characteristic of magnetic diffuse scattering, which
could be modelled by a broad Lorentzian peak centred at the
\(101_\mathrm{mag.}\) peak position, with an isotropic correlation
length at 30 K \(\lambda = 2.8(2)\) Å.

\hypertarget{inelastic-neutron-scattering}{%
\subsection{Inelastic neutron
scattering}\label{inelastic-neutron-scattering}}

\begin{figure*}[!ht]

\centering

\includegraphics[width=17.8cm]{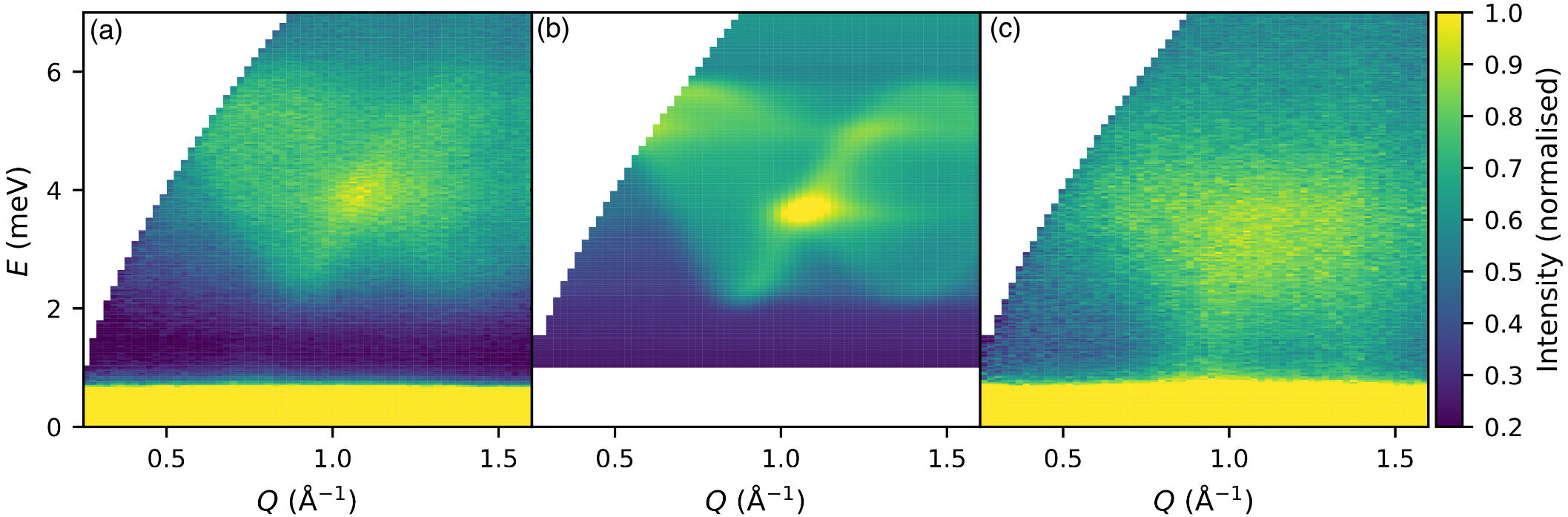}

 \caption{Time-of-flight powder INS spectra of \ce{CrCl2(pym)} with $E_i = 12.14\;$meV measured at (a) 1.7 K and (c) 25 K. (b) LSWT calculated scattering intensity fitted to the 1.7 K data, with parameters $J_1 = -1.13(4)$, $J_2 = 0.10(2)$, $J_3 = 0.01(1)$ and $D =-0.11(1)$ meV. Hamiltonian described in Eqn.~\ref{eq:Hamiltonian}.}

\label{fig:inel}

\end{figure*}

To measure the parameters of the magnetic Hamiltonian and search for
signatures of low-dimensional magnetism, we collected INS spectra on the
same powder sample of \ce{CrCl2(pym)} at 1.7 and 25 K using the LET
spectrometer at ISIS, using rep-rate multiplication to measure at
multiple \(E_i\) simultaneously (\(E_i=12.14, \, 3.70, \, 1.77\) meV).
The spectra collected at 1.7 K show a clear excitation centred at
\(\Delta E = 4.1(2)\;\)meV with an energy gap of \(2.2(1)\;\)meV (Fig.
\ref{fig:inel}a) despite the presence of an elevated background due to
the incoherent \ce{^1H} scattering. The intensity of this feature
rapidly falls with increasing \(Q\), until it is masked by phonons,
indicating this excitation is magnetic in origin. We were able to
quantitatively fit these data using linear spin wave theory (LSWT) (Fig.
\ref{fig:inel}b) as implemented by the SpinW software
package,\citep{tothLinearSpinWave2015} using the following magnetic
Hamiltonian,

\begin{equation}\mathcal{H} = \sum_{\langle ij \rangle} -J_{ij}\mathbf{S}_i \cdot \mathbf{S}_j + \sum_i D(S_i^z)^2, \label{eq:Hamiltonian}\end{equation}

comprising Heisenberg exchange, \(J_{ij}\), for the three nearest
neighbours (\emph{i.e.} along the \ce{CrCl2} through the pym ligand and
between layers) and a single ion anisotropy, \(D\) (Fig.
\ref{fig:pndmag}b). We began by estimating the approximate values for
each of \(J_1\), \(J_2\), \(J_3\) and \(D\) using our bulk magnetic
measurements and extrapolating from analogous
compounds.\citep{stoneQuasionedimensionalSpinWaves2013} These initial
parameters were then optimised using least square requirements of the
calculated spectrum, including a refined multiplicative scale factor and
a background linear in both \(Q\) and \(\Delta E\), against the
experiment data which gave \(J_1=-1.13(4)\) meV, \(J_2=0.10(2)\) meV,
\(0<J_3<0.01(1)\) meV and \(D=-0.15(3)\) meV. The value of D was
corrected for kinematical
consistency\citep{balucaniKinematicConsistencyAnisotropic1979}, as by
default SpinW uses the inconsistent
\(D’ = D[1-\frac{1}{2S}] = \frac{3}{4}D\). A grid search was undertaken
to confirm this as a unique solution. Our experimental spectra were
consistent with a negligible value for \(J_3\), however the ground state
determined by PND indicates that \(J_3\) must be ferromagnetic. The
ratio of \(J_1/J_2 = 11(2)\) indicates that the magnetic interactions in
this materials are primarily one-dimensional. We therefore decided to
investigate the spectrum of \ce{CrCl2(pym)} in the short-range ordered
regime to search for coherent excitations (Fig. \ref{fig:inel}c). Energy
cuts, integrated over momentum transfer, \(0.76<Q<1.84 \;\)Å\(^{-1}\),
showed no clear evidence of a gap in the paramagnetic regime, for both
\(E_i = 12.14\) meV and \(E_i = 3.70\) meV, suggesting this material is
not within the Haldane phase (Fig. S3b), although the comparatively high
temperature compared to the expected gap size, \(T/\Delta = 25\) will
make this challenging.

\hypertarget{density-functional-theory}{%
\subsection{Density-functional theory}\label{density-functional-theory}}

\begin{figure}
\hypertarget{fig:Supercell_Structure_D}{%
\centering
\includegraphics{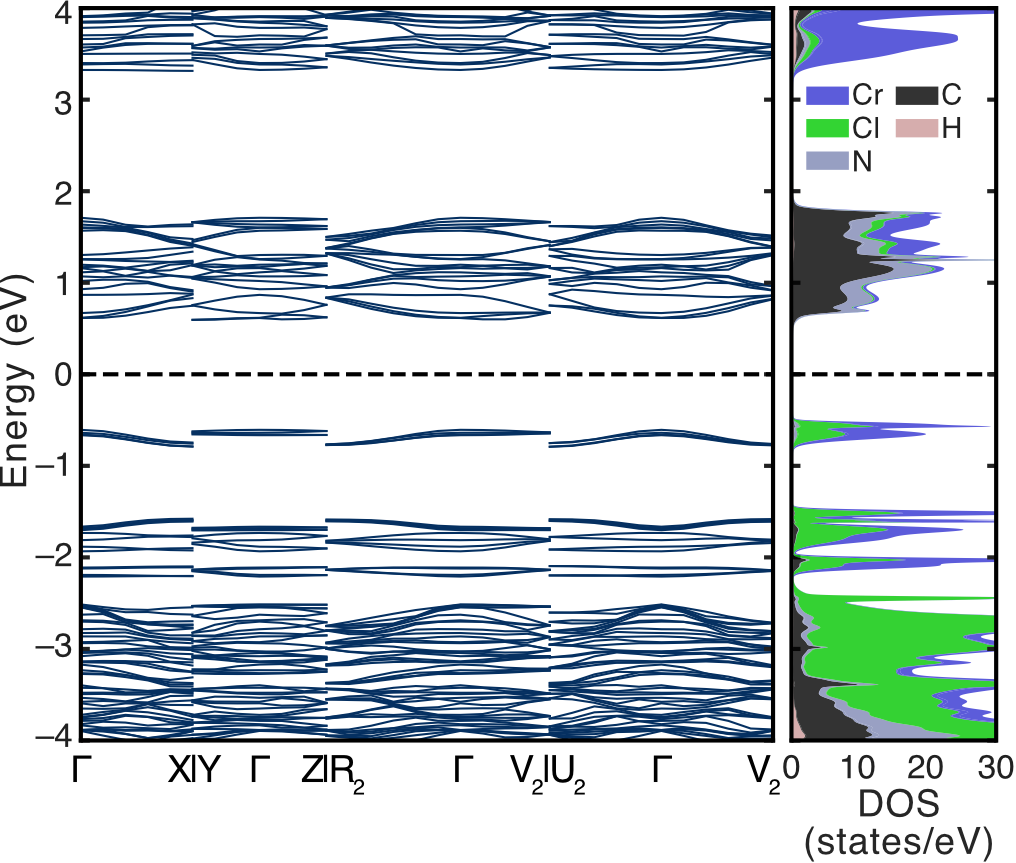}
\caption{Electronic band structure and projected density of states of
the \(2\times2\times1\) supercell using CASTEP and the PBE+U+MBD*
(\(U_\mathrm{eff}=3\;\)eV) functional. The energy zero has been set at
the Fermi energy and is shown by the dashed line. The projected density
of states has been decomposed by
element.}\label{fig:Supercell_Structure_D}
}
\end{figure}

To understand the origin of the observed low-dimensional interactions we
carried out collinear spin-polarized plane-wave density-functional
theory (DFT) calculations, by exploring the electronic structure of the
DFT ground-state spin configuration and calculating the exchange
energies using the broken symmetry
approach.\citep{ciofiniDFTCalculationsMolecular2003} We first optimised
the geometry of the experimental structure using the PBE functional
along with a many-body semi-empirical dispersion correction
(MBD*)\citep{tkatchenkoAccurateEfficientMethod2012} to describe the weak
van der Waals forces between the
layers.\citep{formalikBenchmarkingGGADensity2018} We found that this
structure was both too dense, with a unit-cell volume of 297.68
\AA\(^3\), 4.8\% smaller than experimental value of 312.75 \AA\(^3\),
and lacked the JT distortion characteristic of Cr(II). We therefore
included an effective Coulomb on-site energy, \(U_{\mathrm{eff}} = U-J\)
where \(U\) is the on-site repulsion and \(J\) the exchange energy, to
account for the overly delocalised Cr d-states. A range of values for
\(U_{\mathrm{eff}}\) have been previously explored for Cr, from
\(U_{\mathrm{eff}} = 2.1 \;\)eV to
\(U_{\mathrm{eff}} = 3.5 \;\)eV.\citep{rheeElectronicStructuresOptical2010, wangOxidationEnergiesTransition2006}
We found that \(U_{\mathrm{eff}}=3\;\)eV was able to accurately capture
the physics of this system, and produced a structure with both a JT
distortion and, as a bonus, a volume within +0.2\% of experiment.

Exchange interactions were calculated using a \(2\times2\times1\)
supercell of the optimised structure (\emph{i.e.} containing eight
distinct Cr atoms) decorated with eight distinct magnetic orderings.
Single point energy calculations were then carried out on each
configurations, and these DFT+\(U\) total energies were then fitted to
the Hamiltonian described in Eqn. \ref{eq:Hamiltonian} with \(D = 0\),
\emph{i.e.} the Heisenberg limit. We carried out these calculations
using a series of values of \(U_{\mathrm{eff}}\) to ensure consistency
of behaviour (Fig. S10). For our optimised value of
\(U_{\mathrm{eff}}=3\;\)eV, we obtained a self-consistent set of
superexchange interactions of \(J_1=-2.53(5)\;\)meV,
\(J_2=0.30(5)\;\)meV, \(J_3=-0.09(5)\;\)meV. To test the robustness of
our DFT+U calculations, we performed hybrid calculations using fraction
of Fock exchange as implemented in the HSE
functional\citep{heydHybridFunctionalsBased2003, janeskoScreenedHybridDensity2009, zhuMagneticExchangeInteractions2020}
while maintaining a \(U_\mathrm{eff} = 3\;\)eV. HSE calculations are
computationally expensive due to the calculation of Fock exchange and
require the use of norm-conserving pseudopotentials within CASTEP, which
limited the sampling of the Brillouin zone and our ability to explore
geometry optimisations. Nevertheless, we found that using the HSE
functional comparable exchange interactions \(J_1=-2.39(1)\;\)meV,
\(J_2=0.46(1)\;\)meV, \(J_3=-0.15(1)\;\)meV. These energies are
comparable in magnitude to those found experimentally for
\ce{CrCl2(pym)}, but are notably larger, likely due to the unphysically
large degree of delocalisation.

\begin{figure}
\hypertarget{fig:spin_density}{%
\centering
\includegraphics{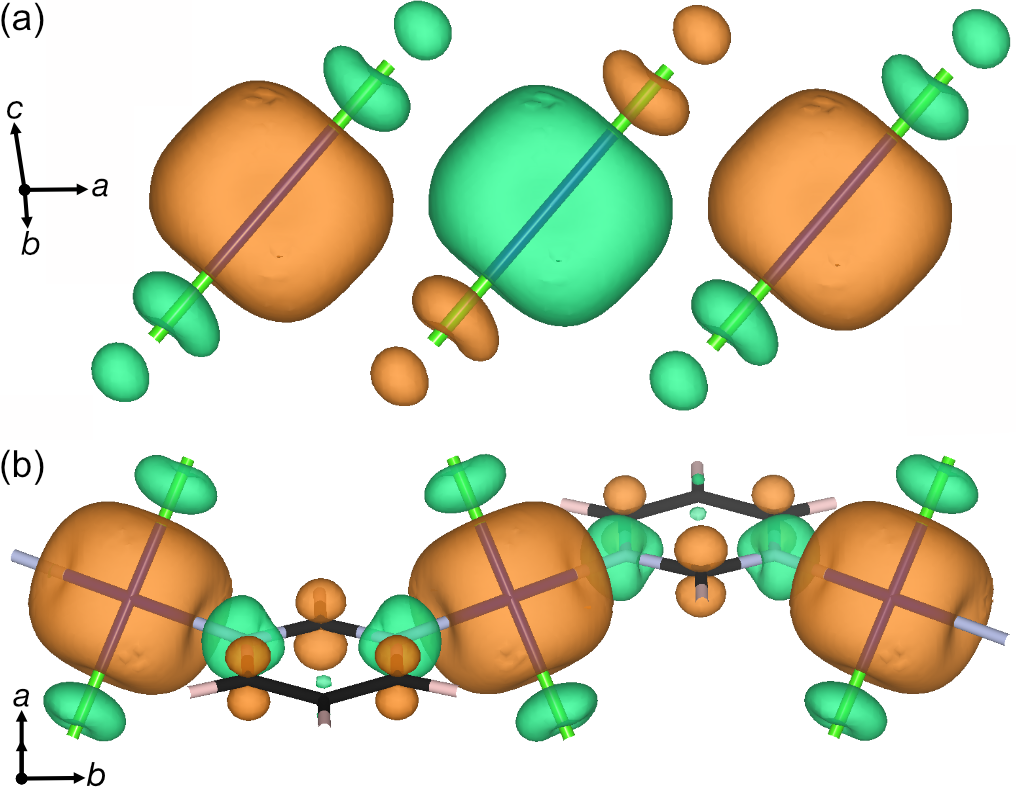}
\caption{Spin density isosurfaces (0.015 e \AA\(^{-3}\)) highlighting
the (a) the Cr--Cl chain and (b) the Cr--pym chain, derived from our
CASTEP PBE+U+MBD* (\(U_\mathrm{eff}=3\;\)eV) and c2x
calculations.\citep{rutterC2xToolVisualisation2018}}\label{fig:spin_density}
}
\end{figure}

Our calculations not only allow us to predict the interaction energies,
but also to explore the electronic structure of this material (Fig.
\ref{fig:Supercell_Structure_D}). The predicted thermal band gap is
approximately \(1.2\) eV, and the projection of the DOS onto local
orbitals shows that the top of the valence band is broadly Cr and Cl
based, whilst the organic linker pym states are the bottom of the
conduction band. This can also be observed in the frontier orbitals,
where the HOMO resembles the Cr \(d_{z^2}\) orbital antibonding with Cl
\(p\) orbitals and the LUMO is an antibonding \(\pi\) molecular orbital
with a single additional node, suggesting that the lowest lying
excitations will be of MLCT character. The spin density is predominantly
around the Cr, however, there is significant density on both Cl and pym
ligands (Fig. \ref{fig:spin_density}). Notably, the spin density on pym
appears to be primarily of \(\pi\) character, and alternates in sign
round the ring (Fig. \ref{fig:spin_density}b).

\hypertarget{discussion}{%
\section{Discussion}\label{discussion}}

Metal N-heterocycle dihalides are a diverse family of MOMs and our study
of \ce{CrCl2(pym)} provides one of the most in-depth investigations of
the magnetic properties of these materials. There are two common
compositions: \ce{MX2L2} and \ce{MX2L}. The monoligand analogues usually
contain linear \ce{MX2} chains, and therefore tend to show primarily 1D
magnetic behaviour, \emph{e.g.} \ce{NiCl2(pyrazine)} consists of
ferromagnetic \ce{NiCl2} chains antiferromagnetically coupled with
\(T_\mathrm{N}=10.2\;\)K,\citep{cortijoModulationMagneticProperties2013}
\ce{CuCl2(pyrazine)} is also a very good example of a 1D magnet with no
order reported down to \(1.8\;\)K, but the strongest interaction in fact
occurs through Cu---pyrazine---Cu bridges (\(J=-28\;\)K), due to the JT
distortion suppressing exchange in the \ce{CuCl2}
chain.\citep{butcherRectangularTwodimensionalAntiferromagnetic2008}
Preliminary studies of the magnetism of pyrimidine analogues,
\ce{MCl2(pym)} M = Mn, Co, Cu, also detect no magnetic order down to 1.8
K although there are weak AFM interactions
present.\citep{zusaiMagnetismPyrimidineBridgedMetal2000} The strong
interactions, particularly occurring through the \ce{CrCl2} chain, and
magnetic order found in \ce{CrCl2(pym)} are therefore in striking
contrast. Additionally, the ferromagnetic exchange we observe occuring
through the pym ligand is relatively uncommon for molecular ligands, for
example, antiferromagnetic interactions are the norm for
pyrazine-bridged
MOMs.\citep{mansonStructuralElectronicMagnetic2011, dossantosExperimentalTheoreticalElectron2016, kubusQuasi2DHeisenbergAntiferromagnets2018, perlepeCrPyrazineOSO2CH32018, natherRationalRouteCoordination2013, wriedtSynthesisCrystalStructure2009, natherThermalDecompositionReactions2003}
This ferromagnetic exchange has been previously observed in pym-bridged
MOMs, \emph{e.g.} \ce{M(NCS)2(pym)2} (\(M=\)Ni and
Co),\citep{lloretTopologySpinPolarization1999, lloretSpinPolarizationFerromagnetism1998, wriedtThermalDecompositionReactions2009}
and has been rationalised by a three-atom \(\pi\)-pathway. Our DFT
calculations give further credence to the importance of this pathway.

The bispyrimidine metal chlorides, \ce{MCl2(pym)2} M = Fe, Co, Ni and
\ce{MBr2(pym)2} M = Co, unlike most materials in this family adopt 3D
chiral diamondoid
structures.\citep{feyerhermWeakFerromagnetismVery2004, hashizumeDichlorobisPyrimidineNCobalt1999, kreitlowPressureDependenceC4N2H4mediated2005}
\ce{MCl2(pym)} all magnetically order with canted AFM structures,
\(T_{\mathrm{N}}\) = 7.4 K, 4.7 K and 16.3 K for M = Fe, Co and Ni
respectively, likely arising from the interplay between the
superexchange interactions and the significant single-ion anisotropy,
the principal axes of which are
non-collinear.\citep{feyerhermWeakFerromagnetismVery2004} Bulk
susceptibility studies have shown enhancement of \(T_\mathrm{N}\) at
moderate pressure (\(\Delta T_\mathrm{N} / T_\mathrm{N} =15\)\% at 0.7
GPa),\citep{kreitlowPressureDependenceC4N2H4mediated2005} which suggests
that high pressure investigations of \ce{Cr}-based MOMs may also uncover
pressure-switchable magnetic
functionality.\citep{wehingerGiantPressureDependence2018}

The presence of a JT distortion is strong evidence of \ce{Cr^{2+}},
which stands in contrast to the related \ce{CrCl2(pyrazine)2}, in which
\ce{Cr^{2+}} spontaneously reduces the ligated pyrazine to a radical
anion, and thereby dramatically enhancing its conductivity and magnetic
superexchange.\citep{pedersenFormationLayeredConductive2018} The
sensitivity of this metal-ligand redox to the coordination sphere is
shown by \ce{Cr(OSO2CH3)2(pyrazine)2}, in which Cr remains as
\ce{Cr^{2+}} with a JT distortion.\citep{perlepeCrPyrazineOSO2CH32018}
Studies of molecular complexes have shown this non-innocent behaviour is
favoured by a strong ligand-field environment and a low energy ligand
LUMO,\citep{scarboroughScrutinizingLowSpinCr2012} and is consistent with
the observed innocence of \ce{CrCl2(pym)}, which has both weaker ligand
field than \ce{CrCl2(pyz)2} and a higher energy ligand LUMO (pyrazine,
\(E_\mathrm{red} = +1.10\;\)V and pym \(E_\mathrm{red} = +0.84\;\)V
vs.~\ce{Li/Li^+}).\citep{assaryReductionPotentialPredictions2014}

Our data clearly show that \ce{CrCl2(pym)} has a conventional Néel AFM
ground state, \(T_{\mathrm{N}}=20.0(3)\) K, but also that there is
significant magnetic low-dimensionality above \(T_\mathrm{N}\).

The frustration parameter,
\(f=\frac{|\theta_{\mathrm{CW}}|}{T_{\mathrm{N}}} = 2.7\), derived from
bulk property measurements hints at suppression of magnetic order. As
the magnetic lattice does not show an obvious mechanism for geometric
frustration, this is likely due to a combination of single-ion
anisotropy and low-dimensionality arising from the large differences in
strength of superexchange in different crystallographic directions.

Additionally, the presence of magnetic diffuse scattering at 30 K not
present at 1.5 K, indicates the presence of short-range magnetic
correlations retained above \(T_{\mathrm{N}}\). Finally, our analysis of
the INS spectra show that the AFM superexchange through the Cr--Cl--Cr
bridge is an order of magnitude larger than all other superexchange
interactions, \(|\frac{J_1}{J_2}| = 11(2)\).

The importance of low-dimensionality can also been seen in the reduction
in the apparent size of the \ce{Cr^{2+}} ordered moment determined via
neutron diffraction. The low-dimensionality of the structure can reduce
the refined moment through disorder, both static short-chain defects and
stacking
faults\citep{winkelmannMagneticOrderingLow1995, hirakawaMagneticNeutronScattering1982}
and dynamic zero-point
fluctuations\citep{cliffeLowdimensionalQuantumMagnetism2018}.
Additionally, as is common in many metal organic
magnets,\citep{kmetyNoncollinearAntiferromagneticStructure2000} there is
appreciable delocalisation of the spin-density onto the ligands, which
Mulliken analysis of the DFT-derived electron density suggests is
approximately 10\%. These factors in combination explain the substantial
reduction in the ordered moment (approximately one-third) from that
expected moment size, though it is challenging to evaluate their
relative contributions.

Despite this low-dimensionality, our data indicate that like other
\(S=2\) candidate AFM chains, \ce{CrCl2(pym)} does not show clear
Haldane physics. The presence of long range order at
\(T_\mathrm{N}/J_1=1.5\) hinders observations at low temperatures and
the non-negligible single ion anisotropy (\(D=-0.15(3)\) meV,
\(D/J_1 = 0.13(2)\)) is sufficient to suppress the Haldane phase, for
which the critical value is predicted to be
\(D/J_1=0.04\).\citep{schollwockHaldaneGapHidden1995} \ce{CrCl2(pym)} is
therefore comparable to the other identified candidate \(S=2\) spin
chains in both of these
parameters,\citep{leoneNeutronDiffractionStudy2004, stockOneDimensionalMagneticFluctuations2009, birkMagneticPropertiesManganese2011}
including \ce{CrCl2}, \citep{stoneQuasionedimensionalSpinWaves2013}
\ce{MnCl3(bipy)}
\citep{granrothExperimentalEvidenceHaldane1996, granrothNeutronscatteringStudiesAntiferromagnetic2002}
and \ce{CsCrCl3},\citep{itohClassicalPropertiesSpin2002} but none have
shown clear evidence of a gapped inelastic neutron spectrum in the
disordered phase.

The compound \ce{CrCl2(pym)} is most similar to, both structurally and
magnetically, is
\ce{CrCl2},\citep{stoneQuasionedimensionalSpinWaves2013} which also has
quasi-1D antiferromagnetic \ce{CrCl2} chains formed from edge-sharing
octahedra (\(J_1=-1.13(13)\) meV, \(D'=-0.11(2)\) meV). However, closer
examination reveals significant structural differences that make these
magnetic similarities quite surprising. In \ce{CrCl2(pym)} the JT
distortion means every superexchange pathway within the \ce{CrCl2} spin
chain passes through a significantly lengthened bond, whereas in
\ce{CrCl2} the equivalent JT distortion lies out of the spin-chain plane
and so all \ce{Cr-Cl} bonds in the chain are short. Superexchange
through a JT-lengthened pathway is ordinarily weak, as is indeed found
for the direction perpendicular to the \ce{CrCl2} spin chain in
inorganic \ce{CrCl2}, with an order of magntiude weaker exchange
\(J_2=-0.12(7)\) meV.

A second distinction between these two compounds is the potential for
tuning the interactions through substitution of the ligands. Replacing
pyrimidine by a larger bridging ligand may reduce inter-chain exchange,
suppressing long-range order and allowing access to the paramagnetic
\(S=2\) quasi-1D AFM at lower temperatures. For example, in \ce{NiCl2L}
substituting pyrazine for 1,2-bis(4-pyridyl)ethane reduces
\(T_{\mathrm{N}}\) from 10.2 K to 5.6
K.\citep{cortijoModulationMagneticProperties2013} Equally, optimisation
of the octahedral coordination environment can minimise \(D\), for
example in a family of closely related \ce{Ni^{2+}} compounds, matching
of the ligand field strengths reduces the size of the easy-plane
anisotropy by a factor of
4.\citep{mansonEnhancingEasyplaneAnisotropy2020} Our measurements of the
INS data already suggest that the interlayer interactions are not
significant, but delamination of these van der Waals sheets, as
demonstrated for other magnetic metal-organic
nanosheets,\citep{lopez-cabrellesIsoreticularTwodimensionalMagnetic2018}
may provide an alternative route to better magnetic isolation. These
results suggest therefore that bridging \ce{CrCl2} spin chains with
organic ligands may provide promising future candidates for \(S=2\)
Haldane chains.

\hypertarget{conclusion}{%
\section{Conclusion}\label{conclusion}}

We have reported the crystal structure, bulk magnetic properties,
magnetic ground state and magnetic excitations of a new coordination
polymer, \ce{CrCl2(pym)}. We have shown that the oxidation state of
chromium in this compound is \ce{Cr^{2+}}, remaining \(S=2\), unlike
related \ce{CrCl2} derived MOMs which undergo redox to form triplet
\ce{Cr^{3+}}-radical ligand pairs.
\citep{pedersenFormationLayeredConductive2018, scarboroughScrutinizingLowSpinCr2012}
\ce{CrCl2(pym)} is found to be a \(S=2\) quasi-one-dimensional
antiferromagnet, with an order of magnitude separation in energy scales
of superexchange, \(|\frac{J_1}{J_2}| = 11(2)\). However we did not find
clear evidence of the Haldane gap in the disordered phase, suggesting
the small \(J_2\) and \(D\) are sufficient in this compound to either
suppress the \(S=2\) Haldane phase or mask it through the stabilisation
of long range order. The proximity of \ce{CrCl2(pym)} to the Haldane
region of the phase diagram, and the modularity inherent to MOMs suggest
that optimising the magnetic properties of these systems, including both
superexchange\citep{cortijoModulationMagneticProperties2013} and
single-ion-anisotropy, \citep{mansonEnhancingEasyplaneAnisotropy2020} is
a new and promising route to the \(S=2\) Haldane phase.

\hypertarget{experimental}{%
\section{Experimental}\label{experimental}}

\hypertarget{synthesis}{%
\subsection{Synthesis}\label{synthesis}}

Synthesis and handling of \ce{CrCl2(pym)} was performed in a dry Ar or
\ce{N2} atmosphere using a MBraun LABstar glovebox or Schlenk line. The
reaction of \ce{CrCl2} (200 mg, 1.63 mmol; Fisher Scientific, 99.9\%)
and pyrimidine (500 mg, 6.24 mmol; Sigma-Aldrich, \(\geq\) 98.0\%) in 50
mL methanol (MeOH) rapidly precipitates an orange-brown microcrystalline
powder. The \ce{CrCl2(pym)} product was then dried \emph{in vacuo}
giving a ca. 90\% total yield. The measured (calculated) elemental
composition was: C, 23.45\% (23.67\%); H, 1.99\% (2.40\%); N, 12.94\%
(13.80\%). This procedure, with quantities scaled up (\ce{CrCl2}, 3.0 g;
pyrimidine, 4.0 g; MeOH, 300 mL), was used to synthesise the sample used
for neutron-scattering measurements. Crystals of sufficient size for
X-ray diffraction studies (\(127 \times 46 \times 26\; \mu\)m) were
grown by vapour diffusion of pyrimidine (100 mg, 1.25 mmol) into a
concentrated solution of \ce{CrCl2} in 1 mL MeOH (10 mg, 0.08 mmol).

\hypertarget{powder-x-ray-diffraction}{%
\subsection{Powder X-ray diffraction}\label{powder-x-ray-diffraction}}

PXRD data were collected using a PANalytical X'Pert Pro diffractometer
equipped with monochromated Cu K\(\alpha\)\(_{1}\) radiation
(\(\lambda= 1.5406\) Å). The tube voltage and current were 40 kV and 40
mA, respectively. Scans were performed from 2\(^{\circ}\) to
60\(^{\circ}\) on a zero background silicon crystal plate. Peak fitting,
Pawley and Rietveld refinement were performed using Topas Academic
v6.\citep{coelhoTOPASTOPASAcademicOptimization2018}

\hypertarget{single-crystal-x-ray-diffraction}{%
\subsection{Single crystal X-ray
diffraction}\label{single-crystal-x-ray-diffraction}}

A diffraction-quality single crystal of \ce{CrCl2(pym)} was mounted on a
polymer-tipped MiTeGen MicroMountTM using Fomblin (YR-1800
perfluoropolyether oil). The sample was cooled rapidly to 120 K in a
stream of cold N\(_{2}\) gas, using a Oxford Cryosystems open flow
cryostat. Diffraction data were collected on an Oxford Diffraction
GV1000 (AtlasS2 CCD area detector, mirror-monochromated Cu-K\(\alpha\)
radiation source; \(\lambda = 1.54184\) Å, \(\omega\) scans). Cell
parameters were refined from the observed positions of all strong
reflections and absorption corrections were applied using a Gaussian
numerical method with beam profile correction (CrysAlisPro). The
structure was solved and refined in
Olex2\citep{dolomanovOLEX2CompleteStructure2009} using
SHELXT\citep{sheldrickSHELXTIntegratedSpacegroup2015} and
SHELXL\citep{sheldrickCrystalStructureRefinement2015}, respectively.

\hypertarget{magnetic-susceptibility-1}{%
\subsection{Magnetic susceptibility}\label{magnetic-susceptibility-1}}

Magnetic property measurements were carried out on a Quantum Design MPMS
superconducting quantum interference device (SQUID). A polycrystalline
sample of \ce{CrCl2(pym)} (26.6 mg) was immobilised in eicosane (44.5
mg) and sealed in a low-paramagnetic-impurity borosilicate glass ampoule
under vacuum. Magnetic susceptibility measurements were performed under
field cooled (FC) and zero-field cooled (ZFC) conditions in a 0.01 T
\emph{dc} field from 2 K to 300 K. Isothermal magnetisation measurements
were performed at 2 K from 0 T to 5 T to --5 T to 0 T. Data were
corrected for the diamagnetism of the sample using Pascal's
constants.\citep{bainDiamagneticCorrectionsPascal2008}

\hypertarget{heat-capacity-1}{%
\subsection{Heat capacity}\label{heat-capacity-1}}

Heat-capacity measurements were carried out on a 4.2 mg pellet of
\ce{CrCl2(pym)} and silver powder (50 wt. \%), using a Quantum Design
Dynacool Physical Property Measurement system (PPMS), between 2 and 60
K. Apiezon N grease was used to ensure good thermal contact.
Contributions to the heat capacity due to Apiezon N were measuerd
separately and subtracted, contributions due to silver were subtracted
using tabulated values.\citep{smithLowTemperaturePropertiesSilver1995}

\hypertarget{powder-neutron-diffraction}{%
\subsection{Powder neutron
diffraction}\label{powder-neutron-diffraction}}

PND measurements were carried out on the D1B neutron diffractometer at
Institut Laue-Langevin, Grenoble, France. Measurements were collected at
1.5 K and 30 K with \(\lambda = 2.52\) Å between \(0.77^{\circ}\) and
\(128.67^{\circ}\) with steps of \(0.1^{\circ}\). The nuclear structure
determined from single crystal X-ray diffraction was Rietveld refined
against neutron diffraction data to evaluate phase purity. Due to the
low intensity of magnetic reflections, the magnetic structure was
determined by refinement against data from which background and nuclear
Bragg peaks were removed by subtraction of data collected at 30 K from
those collected at 1.5 K. The magnetic Bragg peaks were indexed to
determine the magnetic propagation vector and then the allowed magnetic
irreducible representations were determined using symmetry-mode analysis
on the ISODISTORT software.\citep{campbellISODISPLACEWebbasedTool2006}
Using the scale factor determined from Rietveld refinement of the
nuclear structure against data at 30 K, and peak parameters determined
from Pawley refinement of the nuclear structure against data at 30 K,
the direction and magnitude of the ordered moment for the subtracted
dataset were refined using TOPAS-ACADEMIC
6.0.\citep{coelhoTOPASTOPASAcademicOptimization2018}

\hypertarget{inelastic-neutron-scattering-1}{%
\subsection{Inelastic neutron
scattering}\label{inelastic-neutron-scattering-1}}

Inelastic neutron scattering (INS) measurements were performed on the
LET time-of-flight direct geometry spectrometer at
ISIS.\citep{bewleyLETColdNeutron2011} The sample (4 g) was contained in
a thin aluminum can of diameter 15 mm and height 45 mm and cooled in a
helium cryostat. The data were collected at 1.7 K and 25 K, for 10 h and
7 h respectively, with \(E_i = 12.14 \;\)meV using the rep-rate
multiplication
method.\citep{russinaFirstImplementationRepetition2009, russinaImplementationRepetitionRate2010}
The data were reduced using the Mantid-Plot software
package.\citep{arnoldMantidDataAnalysis2014} The raw data were corrected
for detector efficiency and time independent background following
standard procedures.\citep{windsorPulsedNeutronScattering1981}

\hypertarget{density-functional-theory-1}{%
\subsection{Density-functional
theory}\label{density-functional-theory-1}}

Plane-wave density-functional theory calculations were performed using
version 19.1 of the CASTEP code.\citep{clarkFirstPrinciplesMethods2005}
The Brillouin zone was integrated using a Monkhorst-Pack grid of
\(k\)-points, finer than \(2\pi \times 0.05\)
\AA\(^{-1}\)~spacing.\citep{monkhorstSpecialPointsBrillouinzone1976} A
Gaussian smearing scheme with a smearing width of 0.20 eV was used
during the electronic minimisation process. Vanderbilt ultrasoft
pseudo-potentials were used for computational efficiency (Tab.
S3).\citep{vanderbiltSoftSelfconsistentPseudopotentials1990} The basis
set included plane-waves up to an associated kinetic energy of
1100\(\;\)eV. Geometry optimisations converged until resultant forces
were less than \(0.05\) eV/\AA. The OptaDOS post-processing code was
used to integrate individual Kohn-Sham eigenvalues into an electronic
density of states,\citep{morrisOptaDOSToolObtaining2014} and the Matador
high-throughput environment was used to obtain electronic band structure
and density of states plots.\citep{evansMatadorPythonLibrary2020}

\hypertarget{supporting-information}{%
\section{Supporting Information}\label{supporting-information}}

Information on single-crystal and powder X-ray diffraction, additional
inelastic neutron scattering data, powder X-ray and neutron diffraction
data, isothermal magnetisation measurements, magnetic susceptibility
analysis, X-ray photoelectron spectroscopy, transmission electron
micrographs and additional details of DFT calculations (PDF).

CCDC 2213061, crystallographic data (CIF).

Magnetic structure (mCIF).

Additional research data for this Article may be accessed at no charge
and under CC-BY license at the University of Nottingham Research Data
Management Repository https://doi.org/10.17639/nott.7257.

Inelastic neutron scattering data measured at ISIS Neutron and Muon
Source is available at https://doi.org/10.5286/ISIS.E.RB2090119.

CCDC 2213061 contains the supplementary crystallographic data for this
paper. These data can be obtained free of charge via
www.ccdc.cam.ac.uk/data\_request/cif, or by emailing
data\_request@ccdc.cam.ac.uk, or by contacting The Cambridge
Crystallographic Data Centre, 12 Union Road, Cambridge CB2 1EZ, U.K.;
fax: + 44 1223 336033.

\hypertarget{toc}{%
\section{TOC}\label{toc}}

\begin{figure}
\centering
\includegraphics{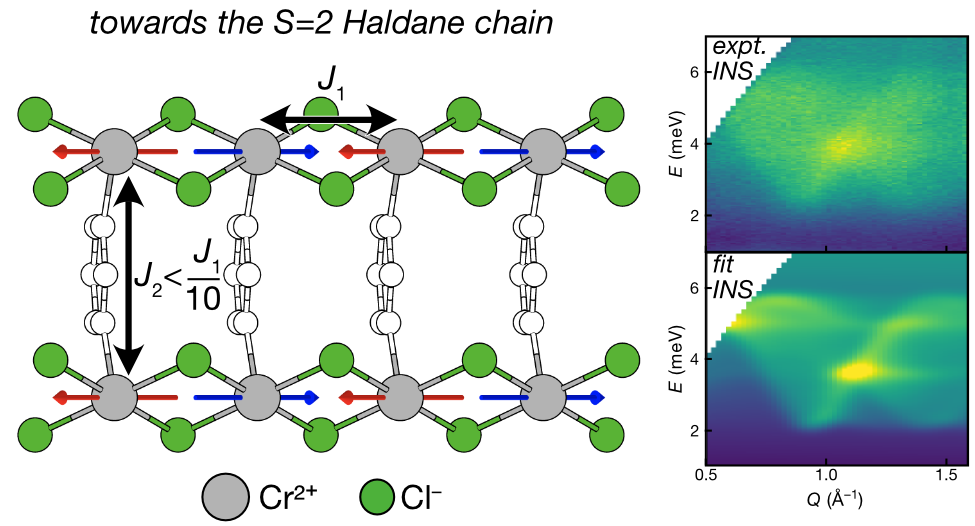}
\caption{TOC Figure}
\end{figure}

\begin{acknowledgement}
J.P. and M.J.C. acknowledge the School of Chemistry, University of
Nottingham for support from the Hobday bequest. A.J.M. acknowledges
funding from EPSRC (EP/P003532/1). The authors acknowledge networking
support via the EPSRC Collaborative Computational Projects, CCP9
(EP/M022595/1) and CCP-NC (EP/T026642/1). Computing resources were
provided by the Sulis HPC service (EP/T022108/1). We acknowledge the ILL
for beamtime under proposal EASY-778. We acknowledge ISIS for beamtime
under proposal RB2090119. J.P. acknowledges Jesum Alves Fernandes for
training and assistance with XPS analysis. J.P. acknowledges Benjamin
Weare for TEM analysis. Heat Capacity measurements were performed using
the Advanced Materials Characterisation Suite, funded by EPSRC Strategic
Equipment Grant EP/M000524/1. J.P. and M.J.C. acknowledge Siân Dutton
and Joseph Paddison for useful discussions.
\end{acknowledgement}

\bibliography{./DefaultLibrary-JP-230109.bib}

\end{document}